# Refactoring of a Database

Ayeesha D'Sousa
Computer Engineering Dept , TSEC ,
Thadomal Shahani Engineering College, Bandra West,
Mumbai 400 050 , India
d.ayeesha@gmail.com

Shalini Bhatia
Computer Engineering Dept , TSEC ,
Thadomal Shahani Engineering College, Bandra West,
Mumbai 400 050 , India
shalini.tsec@gmail.com

*Abstract*—**The technique of database refactoring is all about applying disciplined and controlled techniques to change an existing database schema. The problem is to successfully create a Database Refactoring Framework for databases. This paper concentrates on the feasibility of adapting this concept to work as a generic template. To retain the constraints regardless of the modifications to the metadata, the paper proposes a MetaData Manipulation Tool to facilitate change. The tool adopts a Template Design Pattern to make it database independent. The paper presents a drawback of using java for constraint extraction and proposes an alternative.**

## I. Introduction

Traditional Database Design Approach (TDD) of designing a database schema thinks of Database Design as a phase, which creates a nearly complete set of logical and physical data models, and typically completes before you begin a construction of a project. Initially Database Designers and Software Developers were of the belief that, for a project, once a database is designed, it should not be structurally modified. In fact, it was thought to be impossible. However, requirements change as a project progresses.

TDD has a mechanism for handling and implementing change called Control Change Management. The mechanism works perfectly well when the process is neither over- nor under-managed, but is properly resourced and applied intelligently. Traditional databases are initially well designed. The following two scenarios can result in databases schemas that are very difficult to evolve. One, the change process is made cumbersome to the point where it is unworkable. This results in changes being made very slowly; in practice, too slowly to be effective. Two, Ill-managed changes are rapidly and un-intelligently applied to the database. These cause the structure to degrade over time, rendering it more and more difficult to change the schema. The Traditional Approach does not have the ability to handle a "constantly evolving" Database Schema and has hence "tried to mange change".

Evolutionary Database Design looks at Database Schema Design as an ongoing process interleaved with construction, testing and sometimes even delivery. This is the contrast between TDD and Evolutionary Design in a database.

In Evolutionary Design, on one hand, you had a static database design that could not be frequently changed and on the other hand, you had a constantly evolving, dynamic code. This severely limited the freedom that evolutionary methodologies strived for. Evolutionary DBAs needed to adopt techniques like Refactoring, which enabled them to work in such a regularly changing environment. Small iterative refactoring allow an evolutionary DBA to avoid the mistake of a big upfront design and instead evolve the schema along with the application as they gradually gain a better understanding of the customer requirements

The whole crunch came down to the point that there wasn't any existing guidelines or framework that suggested methods or even assisted in the techniques of database refactoring. Scott Ambler and Pramod Sadalage addressed the problem of a refactoring framework in [1] [2]. This paper concentrates on the feasibility of adapting this concept to work as a generic template.

### *A. Problem Definition*

Database Refactoring is "the act of making a series of simple transformations to a database schema that improves its design while retaining both its behavioral and informational semantics." Each transformation is called 'a refactoring'. The advantage is that you are slowly but constantly improving the quality of the database design, thus over time making the database design easier to understand and hence improving the overall productivity.

An example: Foreign Key Constraints ensure the validity of the data at the cost of the constraint being updated each time that the source data is updated. For a content-Retrieval system, ensuring Foreign Key Constraints at database level increases the response time. The fundamental trade off is still between performance vs. quality. Suppose that the performance cost of enforcing referential integrity by the database cannot be sustained by the database anymore. The problem is that the existing design is not the best possible design. The solution is to refactor the database schema. One solution is to shift the data integrity enforcement to the external application. Applying the Drop Foreign Key Constraint Refactoring ensures that the enforcement of the data dependency is no longer at the database level.

Thus, well thought about changes can be rapidly made by implementing various Refactorings. Refactoring can be broadly classified into [1]:

1) Structural Refactoring (change to the definition of one or more tables or views e.g.: DROP/MERGE/ RENAME columns/table/views)
2) Data Quality Refactoring (change to improve data quality e.g.: MAKE COLUMN NON-NULLABLE, DROP COLUMN CONSTRAINTS)
3) Referential Integrity Refactoring (changes that enforce Referential Integrity),
4) Architectural Refactoring (change that improves the manner in which external programs interact with the database),
5) Method Refactoring (change to a stored procedure, function, trigger) and
6) Transformations (change to database schema that changes its semantics)

Even a small change in schema cascades many changes. Knowing what refactoring is available and the correct sequence of applying the refactoring functions is of paramount importance. The paper proposes a Metadata Manipulation Tool that follows database smells, identifies the various processes to refactor, and aids in finally refactoring the database.

The advantages with Metadata Manipulation Tool are:

1) Minimal upfront design & documentation is required.
2) Initially identifying and thinking of major issues avoids rework.
3) Changes are small and are in sync with the development teams who wish to work in an evolutionary manner

## II. DESCRIPTION

Software is also a complex system, in the sense that it evolves over a period. To efficiently develop such software, we need a software process model that can handle evolutionary changes. Many different models have been proposed since the much cited waterfall model. Conventional software models like the linear sequential model proved inadequate. Next were the evolutionary software process models.

While these techniques have grown in use and interest, one of the biggest questions is how to make evolutionary design work for databases. Most people consider that database design is something that absolutely needs up-front planning. Changing the database schema late in the development tends to cause widespread breakages in application software. Furthermore, changing a schema after deployment, results in painful data migration problems.

The Database Designer starts refactoring the database to make it easier to add the new feature, and after the Refactoring is successfully applied, then adds the new feature. The process of refactoring simplifies to the following process: you have a current state and you wish to achieve a desired state. By applying the Refactoring Function with the Business Rules, Database Rules and Constraints as its guard conditions, you can achieve the next stable state.

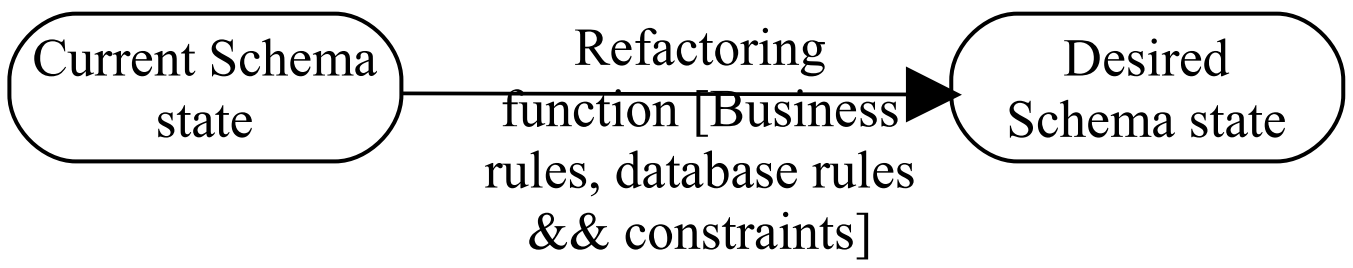

Figure 1: Database Refactoring Function

### A. *Rationale behind Metadata Manipulation Tool*

An area where database metadata comes to play is the dynamic handling of code. Database Metadata can be used to reverse-engineer the whole database and dynamically build desired SQL queries.

Most database store the metadata information as tables themselves. In order to change the structure of the schema, one needs to modify the structure of the database metadata. However, while enforcing ACID properties of the database, the schema too is constrained by these constraints. Hence, to retain the constraints regardless of the modifications to the metadata, a Metadata Manipulation Tool is required to facilitate this change. This tool will implement the Database Refactoring functions. A Metadata Manipulation Tool, will, to an extent, automate and aid in the refactoring process.

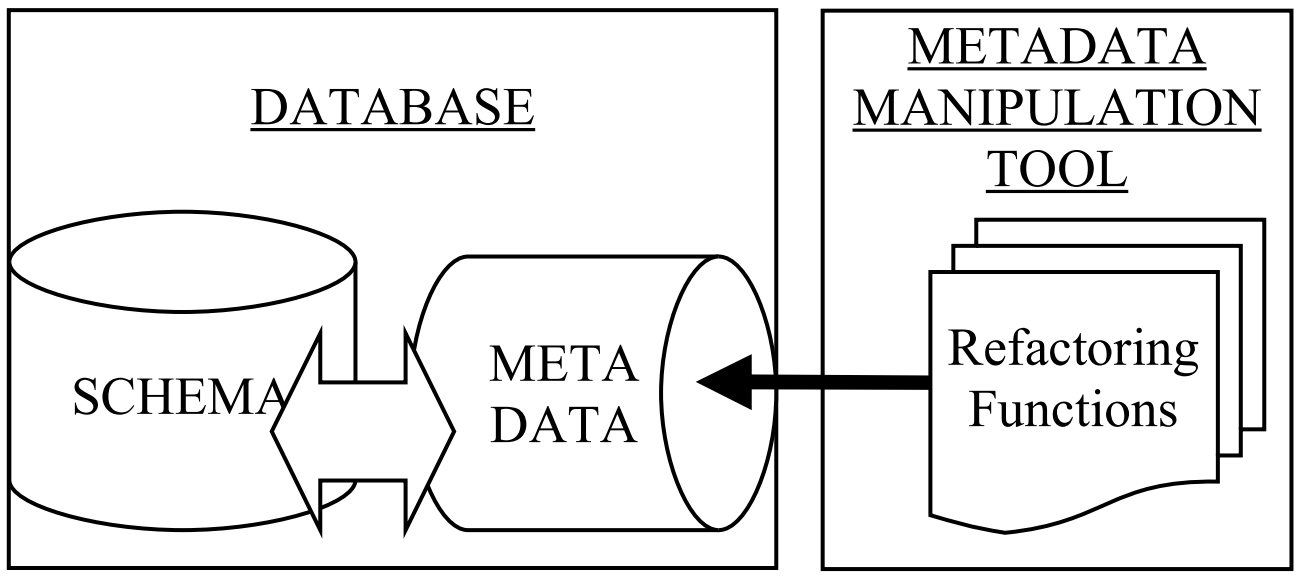

Figure 2. Metadata Manipulation Tool

A Database Designer identifies the issues that need to be resolved. The Database Designer then chooses a solution and plans a Refactoring Function. Typically, the Refactoring Function may work through some or all of the following steps to implement the refactoring. The activity diagram is as follows:

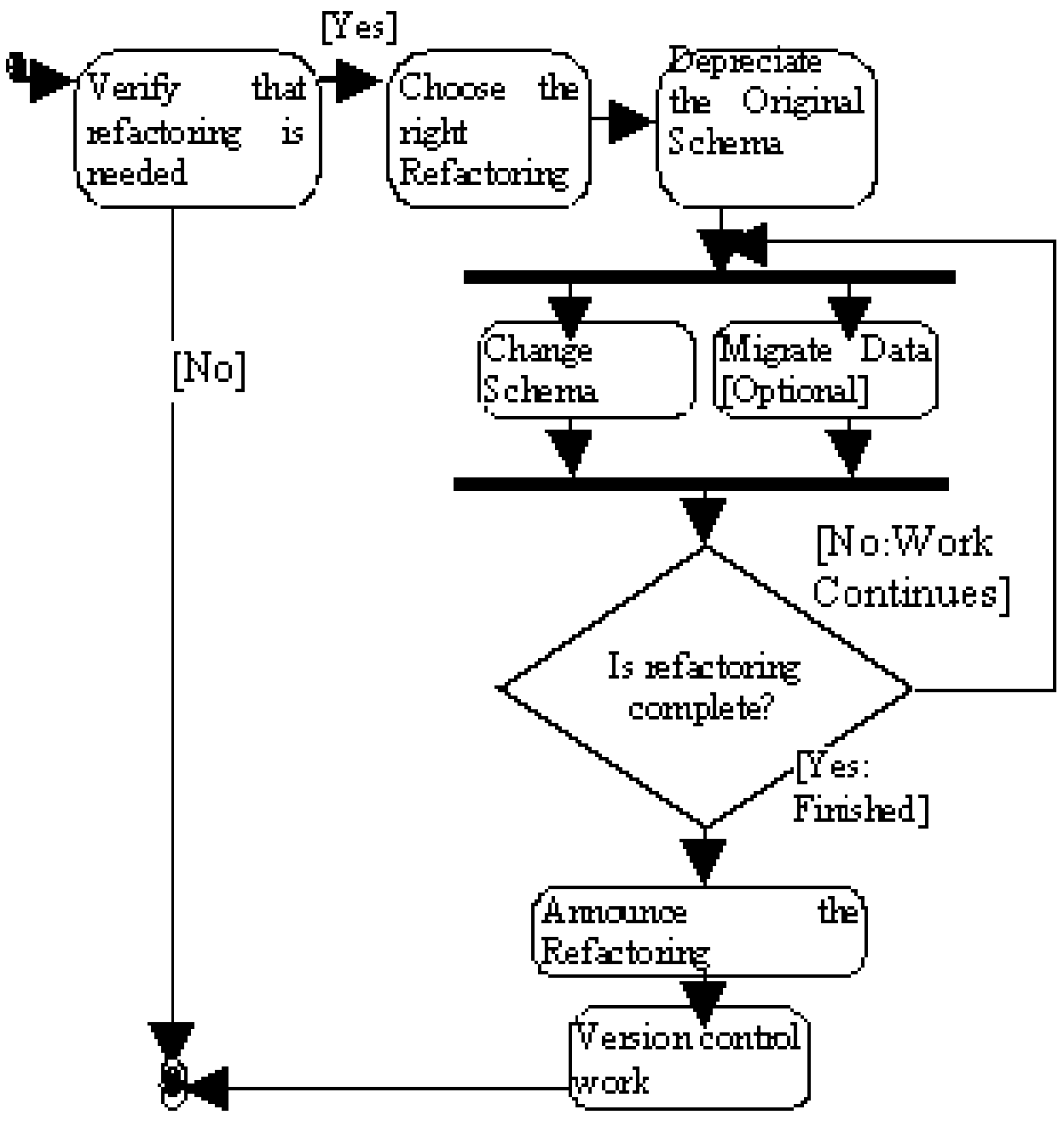

Figure 3. Refactoring Function

B. *The need for Template Design Pattern*

A variance lies in the DDL and DML calls across the databases. Hence, in order to facilitate easy alterations / additions later – a template design pattern is been followed. This allowed the tool to be database independent.

Each refactoring is implemented as an abstract class. A subclass is created for each database, which handles the respective Ddl and DML.

C. *Technique used in Acquiring a Schema Pattern*

The success of the database refactoring framework lies in its ability to accurately extract all the table constraints.

There are two methods to extract schema information:

(1) The Oracle Database Instance. and (2) java.sql.DatabaseMetaData

1) *How Database Objects' Definitions can be extracted from the Oracle Database Instance:*

Previous versions of Oracle provided no specialized ways for DDL extraction. The only way was (a) to execute some SQL-statements on the existing DBA_VIEWS or (b) simply export database schema and parse it [3]. These methods have both limitations and disadvantages. They are: (a) Prior coding techniques / scenarios did not have any modeling tools that were able to develop structures to recreate a database schema or object. One needed to devise SQL scripts that would extract some form of DDL or keep tight control through some form of version control for database structures if one ever wanted to re-create the database from scratch. (b) Export and Import commands for a database schema are difficult to work with for individual objects at times.

Oracle 9i (and higher) has a powerful package - DBMS_METADATA. Functions from DBMS_ METADATA package provide an easy way to get objects' definitions either in XML representation or in DDL. The main function is get_ddl().

Syntax [4]

```
DBMS_METADATA.GET_DDL (
object_type    IN VARCHAR2,
name           IN VARCHAR2,
schema         IN VARCHAR2 DEFAULT NULL,
version        IN VARCHAR2 DEFAULT 'COMPATIBLE',
model          IN VARCHAR2 DEFAULT 'ORACLE',
transform      IN VARCHAR2 DEFAULT 'DDL')
RETURN CLOB;
```

Default Syntax

SELECT DBMS_METADATA.GET_DDL (<OBJECT_TYPE>, <OBJECT_NAME>, <SCHEMA_NAME>) FROM DUAL;

*Object_Type* - PROCEDURE, TABLE, INDEX, CONSTRAINT, etc.

Kindly refer [5] for the Return Values, exceptions, security model and further examples.

2) *Database Objects' Definitions Can Be Extracted From Java (JDBC) : java.sql.DatabaseMetaData, java.sql.ResultSetMetaData [6] [7] [8]:*

JDBC provides a low-level interface called DatabaseMetaData. Most of JDBC's metadata consists of information about one of two things:

1) java.sql.DatabaseMetaData (database metadata information)

The java.sql.Database MetaData interface provides methods for retrieving various metadata associated with a database. Table I offers a partial listing of these methods.

2) java.sql.ResultSetMetaData (metadata information about a ResultSet object)

Example:

getTables (CATALOG, SCHEMA, TABLENAMES, COLUMNNAMES)
getTables(NULL,NULL,TABLENAMES,COLUMNNAMES)

TABLE 1. *SOME DATABASEMETADATA METHODS*

| *Method Name* | *Description* |
|---|---|
| getCatalogs() | To get the name of databases For Oracle, use getSchemas(); MySQL, use getCatalogs(). |
| getSchemas() | Retrieves the schema names (as a ResultSet object) available in this database. |
| getTables(catalog, schema, tableNames, columnNames) | Returns table names for all tables matching tableNames and all columns matching columnNames. |
| getColumns(catalog, schema, tableNames ,columnNames) | Returns table column names for all tables matching tableNames and all columns matching columnNames. |
| getPrimaryKeys(catalog,schema, tableName) | Retrieves a description of the given table's primary key columns. |
| getDriverName() | Gets the name of the database driver you are connected to. |

In the case of methods that return a ResultSet object, either a ResultSet object (which may be empty) is returned or a SQLException is thrown.

Refer [8] for more details.

# III. IMPLEMENTATION

The vital issues of refactoring are: 1] Correct parsing of column level and table level constraints 2] Retrieving of the table schema using the packages described in Section II.C

## *A. Parsing of the Column Level and Table Level Constraint*

Define To test the refactoring, the following are the constraint combinations for the tables in the database.

1. Default table definitions of Employee and Department.
2. Primary Key and Foreign Key (FK references a single table).
3. Only column Definitions (No constraint definitions).
4. Composite Primary Key. In addition one column of the key of the composite Primary key is separately a unique key.

   CREATE TABLE SCOTT.PK2TEMP (
   NUMB NUMBER, VALUE NUMBER,
   CONSTRAINT PK_NUMBVALUE PRIMARY KEY (numb,value) ,
   CONSTRAINT UQ_NUMB (NUMB) )

5. Only Unique key constraints
6. FOREIGN KEY defined on a ( UNIQUE + NOT NULL key )

   **CREATE TABLE SCOTT.FK3TEMP (**
   SRNO NUMBER **NOT NULL,**
   CONSTRAINT UQ_SRNO **UNIQUE** (SRNO)**)**

   **CREATE TABLE FK2TEMP (**
   NUMB2 NUMBER, VALUE2 NUMBER, SRNO NUMBER,
   **CONSTRAINT FK_NUMB2VAL2_PK2TEMP_NUMBVL2 FOREIGN KEY (NUMB2, VALUE2) REFERENCES SCOTT.PK2TEMP (NUMB, VALUE),**
   **CONSTRAINT FK2TEMP_FK03 FOREIGN KEY (SRNO) REFERENCES SCOTT.FK3TEMP**

7. **Check Constraints**

   CREATE TABLE SCOTT.TESTNULL (
   NOTNULLVAL NUMBER (3, 0) NOT NULL,
   ACCEPTNULLVAL NUMBER (3, 0),
   DEFVAL NUMBER (3, 0) DEFAULT 49 NOT NULL,
   UNIQVAL NUMBER (3, 0) NOT NULL,
   PKVAL NUMBER (3, 0) NOT NULL,
   CHECKVAL NUMBER (3, 0) NOT NULL,
   **CONSTRAINT CHECKVAL_CH CHECK (checkval > 10),**
   PRIMARY KEY (PKVAL) ,
   CONSTRAINT UNIQVAL_UQ UNIQUE (UNIQVAL) ).

## *B. Handling Version Control*

The tool uses the database itself to handle versioning. It backs up the table and appends a timestamp to the constraint. Any changes made to the constraint timestamp by the user are stored in the following table.

CREATE TABLE NOVCODE_CONSTRAINTS_MODIFIED

( OWNER VARCHAR2(30) ,

CONSTRAINT_NAME VARCHAR2(50) ,

CONSTRAINT_TYPE VARCHAR2(1) ,

TABLE_NAME VARCHAR2(50) ,

R_OWNER VARCHAR2(50) ,

R_CONSTRAINT_NAME VARCHAR2(50) ,

NEW_MODIFICATION_DATE DATE ,

NEW_CONSTRAINT_NAME VARCHAR2(50) ,

NEW_TABLE_NAME VARCHAR2(50)

)

### C. *Refactoring*

The Metadata Manipulation Tool will have templates created for each of the refactoring. The template will guide the DBA through each step of the refactoring by either gathering appropriate input or making suggestions. The template will also hint about each of the major decisions and conditions that must be handled before the refactoring is initiated.

Structural Refactoring

1. Drop Column
2. Drop Table
3. Merge Columns from Single Table
4. Merge Tables
5. Move Column
6. Rename Column

Referential Integrity Refactoring

1. Drop Constraint

Data Quality Refactoring

1. Introduce Default Value
2. Make Column Non-Nullable

Non-Refactoring Transformations

1. Introduce New Column

## IV. RESULTS AND DISCUSSION

The Metadata Manipulation Tool works for the refactoring mentioned in the scope. Oracle 9i's powerful package - DBMS_METADATA's function get_ddl(), successfully replicates any schema. By using java's java.sql.DatabaseMetaData and java.sql.ResultSetMetaData, there is no way to query and obtain the check constraints or the foreign key constraint defined on a Unique and Not Null constraints.

The tool version controls the schema by appending timestamps. It also handles cases when the constraint names exceed the column size defined by individual databases

An Example:

1. Merge Column

It requests for the number of tables with which the merge operation is to be carried out. The only constraint being the first table entered should be the column, which will be modified. It validates the column name within the table name. If true, the function then asks the user to choose between merging and concatenating the columns. For concatenation the user must chose the delimiter. The function then asks the user if he desires a backup of the column and then proceeds with backing up the entire table, which has the modified value.

For merging the user must enter the condition in the Update statement. For concatenation the tool adjusts the length of the column and proceeds by appending the delimiter after each column being merged. See Figure 4 and Appendix A for Implementation.

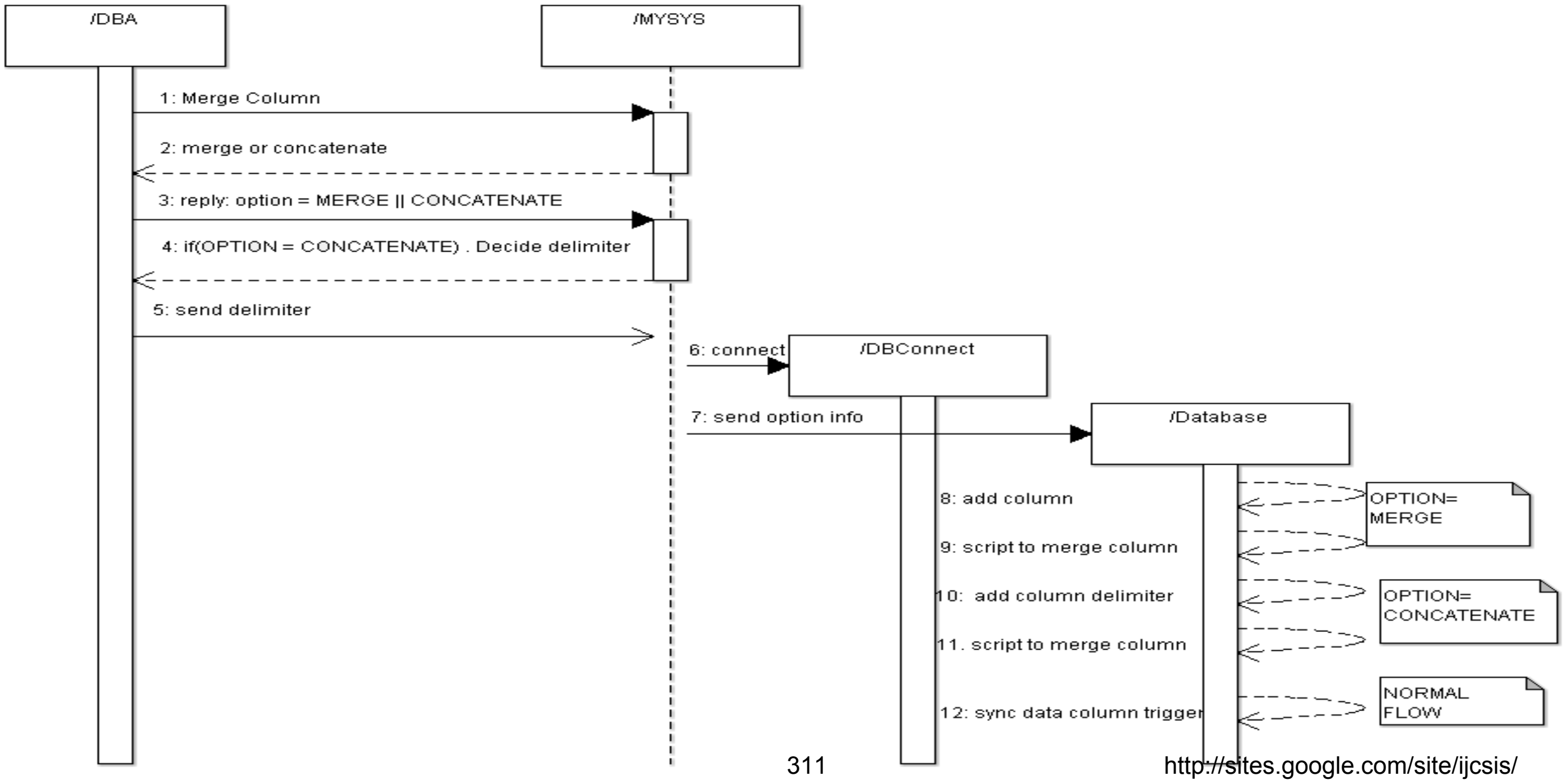

Figure 4: Sequence diagram for Merge Column Refactoring

## V. CONCLUSION

On testing the code on Oracle9i and Mysql – it was realized that the code cannot be made database independent. Hence in order to facilitate easy alterations / additions later – a template design pattern has been followed.

While coding for Oracle 9i – there are two ways to obtain the schema. One method is to use oracle's function DBMS_METADATA.GET_DDL in package DBMS_METADATA and the second method is JDBC's metadata from java.sql. DatabaseMetaData.

The disadvantage of using JDBC's metadata from java.sql.DatabaseMetaData is that there is no method to find the check constraints or a foreign key based on a (unique + not null) key. However all the constraints can be found by using DBMS_METADATA.GET_DDL() in package DBMS_METADATA.

For Oracle9i , currently the scope of the project is limited to refactoring within the same schema or user. The security model of GET_DDL in package DBMS_METADATA prevents access. Nonprivileged users can see the metadata of only their own objects. SYS and users with SELECT_CATALOG_ROLE can see all objects.

For the refactoring that the paper discuss – the algorithms designed works.

## VI. FURTHER WORK

The code does not handle multi-application database environment. This project can be extended to handle and iron out these issues.

## Acknowledgment

Ayeesha Dsousa thanks Prof. Rajesh Patkar and Ashwin Dsousa for their help and technical guidance.

## References

[1] Ambler, S.W. and Sadalage, P.J. (2006). Refactoring Databases: Evolutionary Database Design. *Boston: Addison Wesley*

[2] Ambler, S.W. Test-Driven Development of Relational Databases. *Software, IEEE.* Volume: 24, Issue: 3, May-June 2007, pp. 37-43.

[3] James Koopmann. Tapping into Oracle's Metadata - Part I . *http://www.orafaq.com/node/57 2005-2009*

[4] *Oracle® Database Utilities. 10g Release 1(10.1)*http://www.mcs.csueastbay.edu/support/ *oracle/doc/10.2/appdev.102/b14258/ d_metada.htm#sthref4210 Copyright © 1996, 2003 Oracle Corporation*

[5] Oracle® Database Utilities. 10g Release 1 (10.1) *http://download-west.oracle.com/ docs/cd/B12037_01/server.101/b10825/metadata_api.htm Copyright © 1996, 2003 Oracle Corporation*

[6] Parsian, M. JDBC Metadata, MySQL, and Oracle Recipes: A Problem-Solution Approach (Expert's Voice in Java) (Hardcover). *Apress Academic* . 2006.

[7] Parsian, M. JDBC Metadata, MySQL, and Oracle Recipes: A Problem-Solution Approach (Expert's Voice in Java) (Hardcover). *http://blog.codebeach.com/2008/12/database-metadata-with-jdbc.html* Copyright 2000 - 2009 Code Beach

[8] JavaTM 2 Platform Std. Ed. v1.4.2 http://java.sun.com/j2se/1.4.2/docs/api/java/sql/DatabaseMetaData.html. Copyright 2003

## AUTHORS PROFILE

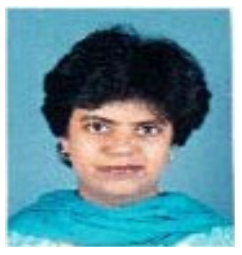

Shalini Bhatia was born on August 08, 1971. She received the B.E. degree in Computer Engineering from Sri Sant Gajanan Maharaj College of Engineering, Amravati University, Shegaon, Maharashtra, India in 1993, M.E. degree in Computer Engineering from Thadomal Shahani Engineering College, Mumbai, Maharashtra, India in 2003.

She has been associated with Thadomal Shahani Engineering College since 1995, where she has worked as Lecturer in Computer Engineering Department from Jan 1995 to Dec 2004 and as Assistant Professor from Dec 2004 to Dec 2009. She has published a number of technical papers in National and International Conferences. She is a member of CSI and SIGAI which is a part of CSI.

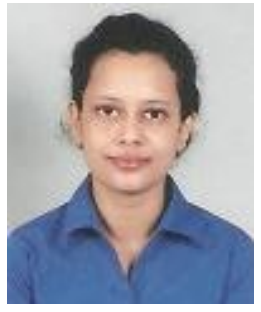

Ayeesha Dsousa was born on July 1, 1980. She received the B.E. degree in Computer Engineering from Sri Sant Gajanan Maharaj College of Engineering, Amravati University, Shegaon, Maharashtra, India in 2002,. She is pursuing M.E. degree in Computer Engineering from Thadomal Shahani Engineering College, Mumbai, Maharashtra, India.. She was working with St.Francis Institute of Technology, Borivli (W) Mumbai from the year 2004 as Lecturer in Information Technology Department. Her research interests include Databases, Advanced Databases and Data warehousing.

## APPENDIX A

**KEY**

**[1] The Metadata Manipulation Tool's prompt to the user**
**[2]** The **user's response**
**[3]** *The output from the* Metadata Manipulation Tool

## RESULTS

11 - Jul - 2009 11:33:41
**Enter Choice for the Database Technology Else press enter to choose Oracle for default**

**Enter the username. Press Enter to use default**

**Enter the password. Press Enter to use default**

*Driver : oracle.jdbc.OracleDriver*
*connect made by : userid scott password tiger*
**Choose Refactoring by entering the number**
Structural Refactoring
1. Drop Column
2. Drop Table
4. Merge Columns : for single table only
5. Merge Tables
Referential Integrity Refactoring
24. Drop Constraint
Data Quality Refactoring
31. Introduce Default Value
32. Make Column Non Nullable
Data Transformations
41. Add New Column
HouseKeeping
91. create tables to test
92. Display constraints on table
93. Display table schema
94. Display table
99. exit
Type :5

5. Merge Tables
**you can merge between 2 tables .The FIRST TABLE that you enter WILL BE BE MODIFIED .**
**enter the 1 Tablename** :EMP24JAN09120206
**enter the 2 Tablename** :suppliers

*Details entered for tables in function to Merge column*

| [0]TABLE_NAME | [1]COLUMN_NAME | [3]TYPE_NAME | [4]COLUMN_SIZE | [5]DECIMAL_DIGITS | [7]COLUMN_DEF | [8]ORDINAL_POSITION | [9]IS_NULLABLE |
|---|---|---|---|---|---|---|---|
| EMP24JAN09120206 | EMPNO | NUMBER | 4 | 0 | null | 1 | NO |
| EMP24JAN09120206 | ENAME | VARCHAR2 | 26 | null | null | 2 | YES |
| EMP24JAN09120206 | JOB | VARCHAR2 | 9 | null | null | 3 | YES |
| EMP24JAN09120206 | MGR | NUMBER | 4 | 0 | null | 4 | YES |
| SUPPLIERS | SUP_ID | NUMBER | 3 | 0 | null | 1 | NO |
| SUPPLIERS | SUP_NAME | VARCHAR2 | 32 | null | null | 2 | YES |
| SUPPLIERS | STREET | VARCHAR2 | 32 | null | null | 3 | YES |

| [0]TABLE_NAME | [1]COLUMN_NAME | [3]TYPE_NAME | [4]COLUMN_SIZE | [5]DECIMAL_DIGITS | [7]COLUMN_DEF | [8]ORDINAL_POSITION | [9]IS_NULLABLE |
|---|---|---|---|---|---|---|---|
| SUPPLIERS | CITY | VARCHAR2 | 32 | null | null | 4 | YES |

**Unless stated otherwise, Press Y for YES, Press N for No.**
**Type the column name that you would like to shift to table EMP24JAN09120206**
street

**Continue entering more column names**
n

*Details entered for tables in function to Merge column*

| [0]TABLENAME | [1]NoOfRows |
|---|---|
| EMP24JAN09120206 | 9 |
| SUPPLIERS | 6 |

*Details entered for tables in function to Merge column*

```
CREATE TABLE SUPPLIERS
   (  SUP_ID NUMBER(3,0),
      SUP_NAME VARCHAR2(32),
      STREET VARCHAR2(32),
      CITY VARCHAR2(32),
      STATE VARCHAR2(3),
       CONSTRAINT SUPP_PK PRIMARY KEY (SUP_ID)
    )
```

*Details entered for tables in function extract constraints*

| [0]Column Name | [1]Row found at | [2] Table Name | [3] Constraint String |
|---|---|---|---|
| SUP_ID | 1 | SUPPLIERS | SUP_ID NUMBER(3,0) CONSTRAINT SUPP_PK PRIMARY KEY (SUP_ID) |
| SUP_NAME | 2 | SUPPLIERS | SUP_NAME VARCHAR2(32) |
| STREET | 3 | SUPPLIERS | STREET VARCHAR2(32) |
| CITY | 4 | SUPPLIERS | CITY VARCHAR2(32) |
| STATE | 5 | SUPPLIERS | STATE VARCHAR2(3) |

*Value of update function is true*
*ALTER table EMP24JAN09120206 ADD STREET VARCHAR2(32)*
**Do you wish to continue with the changes? Press Y for Yes, N for No :** Y

**Choose Refactoring by entering the number**
Structural Refactoring
1. Drop Column
2. Drop Table
4. Merge Columns : for single table only
5. Merge Tables

Referential Integrity Refactoring
24. Drop Constraint

Data Quality Refactoring
31. Introduce Default Value
32. Make Column Non Nullable

Data Transformations
41. Add New Column

HouseKeeping

```
91. create tables to test
92. Display constraints on table
93. Display table schema
94. Display table
99. exit
Type        :93
```

*93. Display table Schema*
*EMP24JAN09120206*

```
Table description
       column names :
```

| | | | | | |
|---|---|---|---|---|---|
| [1] TABLE_CAT | null | null | null | null | **null** |
| [2] TABLE_SCHEM | SCOTT | SCOTT | SCOTT | SCOTT | **SCOTT** |
| [3] TABLE_NAME | EMP24JAN09120206 | EMP24JAN09120206 | EMP24JAN09120206 | EMP24JAN09120206 | **EMP24JAN09120206** |
| [4] COLUMN_NAME | EMPNO | ENAME | JOB | MGR | **STREET** |
| [5] DATA_TYPE | 3 | 12 | 12 | 3 | **12** |
| 6] TYPE_NAME | NUMBER | VARCHAR2 | VARCHAR2 | NUMBER | **VARCHAR2** |
| [7] COLUMN_SIZE | 4 | 26 | 9 | 4 | **32** |
| [8] BUFFER_LENGTH | 0 | 0 | 0 | 0 | **0** |
| [9] DECIMAL_DIGITS | 0 | null | null | 0 | **null** |
| [10] NUM_PREC_RADIX | 10 | 10 | 10 | 10 | **10** |
| [11] NULLABLE | 0 | 1 | 1 | 1 | **1** |
| [12] REMARKS | null | null | null | null | **null** |
| [13] COLUMN_DEF | null | null | null | null | **null** |
| [14] SQL_DATA_TYPE | 0 | 0 | 0 | 0 | **0** |
| [15] SQL_DATETIME_SUB | 0 | 0 | 0 | 0 | **0** |
| [16] CHAR_OCTET_LENGTH | 22 | 26 | 9 | 22 | **32** |
| [17] ORDINAL_POSITION | 1 | 2 | 3 | 4 | **10** |
| [18] IS_NULLABLE | NO | YES | YES | YES | **YES** |